\documentclass[twocolumn,noshowpacs,amsmath,amssymb,nofootinbib]{revtex4}

\usepackage{graphicx}

\begin{document}

%\preprint{}

\title{Field-Theoretic Simulations of Polyelectrolyte Complexation}

\author{Yuri O.\ Popov}
%\email[Email address: ]{yopopov@mrl.ucsb.edu}
\author{Jonghoon Lee}
\author{Glenn H.\ Fredrickson}
\email[Email address: ]{ghf@mrl.ucsb.edu}
\affiliation{Materials Research Laboratory, University of California, Santa Barbara, California 93106, USA}

\date{\today}

%\begin{abstract}
%\end{abstract}

%\pacs{87.14.Gg, 87.15.La, 87.15.Aa}

\maketitle

This viewpoint article is intended as a brief introduction to the emerging subject of field-theoretic simulations (FTS) of charged polymers.  While the direct numerical simulation of field theory models has begun to impact several traditional areas of polymer science, including blends and block copolymers, polyelectrolytes have hitherto not been the subject of field-theoretic simulations.  Here we report on a preliminary FTS study of polyelectrolyte complexation that demonstrates the potential of this novel numerical approach.

Polyelectrolytes are ubiquitous in nature and in applications ranging from personal care products to paints, coatings, and processed foods.  Indeed, practically all biopolymers are polyelectrolytes.  In the application context, the introduction of dissociable groups is one of the most powerful ways to confer water solubility on a polymeric material.  Scientifically, the polymer bound charges, which are compensated by a sea of oppositely charged counterions, produce a coupling between chain conformations and electrostatics that leads to an incredible richness of polyelectrolyte phenomena.  However, this richness comes with a price: charged polymers are among the most difficult polymer systems to study theoretically or to simulate on the computer~\cite{degennes,barrat,holm,forster}.  The explanation lies in the long range nature of Coulomb interactions -- charged segments feel each other at much larger distances than segments in neutral polymer systems.

The situation becomes particularly challenging for the dense polyelectrolyte complexes that are the subject of the present work.  Analytical theories based on assumptions of low concentration or weak interactions break down, and equilibration times in numerical simulations become prohibitively long.  To conduct such simulations, polyelectrolytes are often modeled as coarse-grained chains of charged beads and the counterions are taken to be point particles, usually embedded in an implicit solvent with uniform dielectric properties.  The long-ranged character of Coulomb interactions is problematic for such ``particle-based'' modeling approaches, however, because Ewald sums and other expensive computational techniques are required to evaluate contributions to electrostatic energies and forces that extend beyond the computational cell~\cite{stevens,limbach,frenkel}.  The confluence of this long-range effect with the intrinsically slow kinetics of particle-based simulations at high density and molecular weight produces significant challenges for these methods.

A different strategy to tackle such problems was proposed long ago by S.~F.~Edwards~\cite{edwards}.  The idea is to replace the coordinates and momenta of particles (polymer segments) with \textit{collective} variables, or \textit{fields}.  One can introduce, for example, fields that describe the density of polymer segments and the density of charge.  Moreover, by augmenting these fields with certain \textit{conjugate} fields, such as a chemical potential (conjugate to segment density) and an electrostatic potential (conjugate to charge density), it is possible to \textit{exactly} transform any classical (equilibrium) particle-based model into a statistical field theory~\cite{fredrickson}.  This field-based description is particularly useful for dense polymer systems, such as concentrated solutions and melts, where there is strong overlap among polymers.  In such cases, the \textit{mean-field approximation}, also known as \textit{self-consistent field theory} (SCFT), can be applied with confidence.  Within this approximation, the saddle point of the Hamiltonian dominates the partition function, and the field fluctuations around the saddle point are ignored~\cite{helfand}.  The SCFT method has yielded some remarkable analytical and numerical results for a wide variety of dense equilibrium polymer systems, perhaps most notably in the field of block copolymers~\cite{leibler,matsen}.

Besides its restriction to systems at equilibrium, a major limitation of SCFT is that the assumption of negligible field fluctuations breaks down rapidly as polymers are diluted in a solvent.  This is especially problematic for the study of polyelectrolytes, since they are highly solvated in most applications.  Moreover, polyelectrolytes are characterized by very strong \textit{charge correlations} in addition to the density correlations of neutral polymer solutions.  Evidently the well-tuned machinery of SCFT is not the appropriate tool for investigating solutions of charged macromolecules.

One way to address the limitations of SCFT, while still preserving the field-based collective variable approach, is to return to the exact statistical field theory and to account for the field fluctuation effects.  Analytically, this procedure leads to a so-called ``loop'' expansion, in which systematic corrections to the free energy or other thermodynamic quantities (one-loop, two-loop, etc.) are developed in terms of integrals over field fluctuations about the saddle point of the theory.  Such loop expansions can be further augmented by renormalization techniques in cases of strong fluctuations~\cite{degennes,amit}.  While powerful in the context of \textit{homogeneous phases} of charged and uncharged polymers, these analytical methods can be very difficult to implement in more general inhomogeneous situations where interfaces or mesophases are present.

It has recently been demonstrated that statistical field theory models of polymers can also serve as the basis for computer simulations -- so-called \textit{field-theoretic simulations} (FTS)~\cite{fredreview,fredrickson}.  While similar conceptually to numerical SCFT, field-theoretic simulations aim to numerically sample the statistically important field configurations of the full theory, rather than just the saddle point configuration.  This statistical sampling is problematic because the Hamiltonians of the relevant field theories are complex, rather than strictly real; a manifestation of the famous ``sign problem''.  We have found that a powerful way to circumvent this difficulty is to adopt the \textit{complex Langevin} stochastic procedure~\cite{parisi,ganesan}, which adaptively samples field configurations along nearly constant phase trajectories.  Although FTS is computationally more expensive than SCFT, recent numerical advances have made high resolution FTS feasible for a wide variety of polymer systems~\cite{lennon}.

Polyelectrolytes have been the subject of extensive theoretical and computational research for decades~\cite{degennes,barrat,holm,forster,thunemann,dekruif,shi,borue1,borue2,castelnovo1,castelnovo2,tsonchev,wang,kudlay1,kudlay2,nyrkova,shusharina,liao,dobrynin,zhang}.  Statistical field theory models have played a significant role in these theoretical investigations, and both mean-field and non-mean-field approaches have been employed to gain insights into the structure and thermodynamics of a wide variety of polyelectrolyte systems~\cite{borue1,borue2,castelnovo1,castelnovo2,tsonchev,wang,kudlay1,kudlay2}.  Prior to this work, however, there has been no general numerical tool for simulating a field theory model of polyelectrolytes without the use of simplifying approximations.

The present article describes the first application of FTS to polyelectrolytes, and specifically to a phenomenon that is known to occur in aqueous mixtures of two oppositely charged polymers.  Under appropriate conditions of charge density, solvent quality, and molecular weight, a phase transition can occur in such a system, with two phases being formed: one rich in both polymers and the other consisting of nearly pure solvent~\cite{thunemann,dekruif,dejong,michaele,overbeek,veis1,veis2,veis3,biesheuvel}.  This process is usually referred to as \textit{polyelectrolyte complexation} in the physics community or as \textit{complex coacervation} in the physical chemistry, colloid science, and biological communities.  The resulting polymer-rich \textit{coacervate} phase has two important properties: it is dense yet liquid, and it is charge neutral.

Coacervates and related polyelectrolyte complexes have a variety of important applications.  For example, complexes can be employed as carrier systems for charged macromolecules including protein drugs, enzymes, and DNA.  In such systems, one of the polyelectrolytes serves as a chaperon with properties that can be tailored to assist targeted delivery, while the other represents the macromolecular payload~\cite{thunemann}.  Another industrial use of complex coacervation is in cements, glues, or adhesives, where the coacervate phase can serve as a precursor to the formation of a solidified structural material.  Examples of such coacervate precursors are also encountered in nature; e.g.\ sand-castle worms produce a strong protein-based cement that sets in sea water and is used to construct meter-scale reefs out of sand and shells~\cite{zhao}.  Other existing and potential applications of polyelectrolyte complexes include water purification, DNA sensors~\cite{hong}, and encapsulation of food and pharmaceuticals.

Polyelectrolyte complexation is driven by several competing factors~\cite{thunemann,dekruif,dejong,michaele,overbeek,veis1,veis2,veis3}.  In addition to the direct electrostatic attraction between oppositely charged polyions, there are other factors playing an important role: electrostatic screening by small ions (salt or counterions), excluded volume interactions of polymer backbones mediated by the solvent, small ion translational entropy, and polymer conformational entropy.  In many cases these factors are competitive rather than reinforcing, so it can be difficult to anticipate the conditions under which coacervate phases will form.  For example, excluded volume interactions oppose complexation, while the translational entropy of counterions tends to favor it.

\begin{figure}
\includegraphics{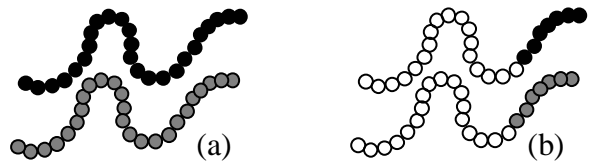}
\caption{\label{one}  Two model polyelectrolyte mixtures.  Both are symmetric, with half of all chains being positively charged (shown in black) and the other half negatively charged (in grey).  (a) The total charge $\pm \sigma N$ is evenly distributed over the entire length of each chain.  (b) The same total charge is concentrated in blocks of $N/4$ segments near one end of each chain.  Uncharged portions are shown in white.}
\end{figure}

To investigate the physics of complex coacervation we have chosen a simple yet fundamental model system [Fig.~\ref{one}(a)].  Specifically, we consider a symmetric polycation-polyanion mixture in an implicit solvent without salt.  Such a system could be realized by mixing a polyacid with a polybase in water.  The symmetric assumption implies that the molecular weights and charge densities of the two types of polymers are the same, and they are combined in equal amounts.  The polymer backbones are modeled as flexible (Gaussian) chains of length (number of statistical segments) $N$, differing only by the sign of the charge per statistical segment (charge density), $\pm \sigma$.  We employ a canonical ensemble in which $n$ polyanions and $n$ polycations are mixed to form a solution of volume $V$.  The chains are assumed to interact by means of electrostatic and excluded volume interactions, characterized by the Bjerrum length $l_B = e^2/\epsilon k_B T$ and the excluded volume parameter $u_0$, respectively.  The solvent dielectric constant is denoted by $\epsilon$, and $e$ is the fundamental unit of charge.  Note that the addition of salt, consideration of asymmetric polyions, or explicit inclusion of the solvent constitute only minor modifications of the formalism, although these variations will not be pursued here.

The thermodynamic properties of the above model can be deduced from a configurational partition function $Z$, which is a sum over all possible states (represented by the coordinates of all polymer segments) of the Boltzmann factor $\exp( - U / k_B T)$.  The energy of each state $U$ can be expressed as a sum of three contributions: Gaussian chain connectivity, excluded volume interactions, and electrostatic (Coulomb) interactions.  Rather than working with this ``particle-based'' statistical mechanical description, we reexpress the partition function in terms of collective field variables~\cite{fredrickson}.  Indeed, $Z$ can be written as a sum (a functional integral) over all \textit{field} configurations:
\begin{equation}
Z = Z_0 \iint Dw D\phi \, \exp\left( - H[w,\phi] \right),
\label{z}
\end{equation}
where $Z_0$ is the partition function of an ideal gas of non-interacting polymers, $w(\mathbf{r})$ is a real chemical potential field conjugate to the segment density, and $\phi(\mathbf{r})$ is a real electrostatic potential field conjugate to the total charge density.  The energy of each state is now replaced by an \textit{effective} Hamiltonian, which for our model is
\[
H[w,\phi] = \frac{1}{2 u_0} \int d^d \mathbf{r} \left[ w(\mathbf{r}) \right]^2 + \frac{1}{2 K_d l_B} \int d^d \mathbf{r} \left| \nabla \phi(\mathbf{r}) \right|^2
\]
\begin{equation}
- \, n \ln Q\left[ i w + i \sigma \phi \right] - n \ln Q\left[ i w - i \sigma \phi \right].
\label{h}
\end{equation}
In this effective Hamiltonian, direct interactions between polymers \textit{decouple}, and the remaining interactions are field-field and field-polymer.  Polymers interact with the fields only via the \textit{single-chain} partition function $Q[\psi]$, where $\psi(\mathbf{r})$ corresponds to the two purely imaginary fields $i w \pm i \sigma \phi$ in Eq~\ref{h}.  This partition function is calculated as a volume integral of a chain propagator, which in turn can be obtained from the solution of a complex diffusion equation in the imaginary field $\psi$~\cite{fredrickson}.  $K_d$ is the coefficient in front of the Green's function of the Laplacian in $d$ dimensions ($4 \pi$ in 3D and $2 \pi$ in 2D).

It is important to note that this field theory is an \textit{exact} reformulation of the original particle-based model.  While the fundamental model is not new, previous field-theoretic approaches to polyelectrolyte complexation~\cite{borue2,castelnovo1,kudlay1} have simplified the field theory by use of weak inhomogeneity expansions.  This simplification is not exact, nor is it desirable or necessary.  We further note that the long-ranged Coulomb interaction in the particle description is replaced by a short-ranged ``square gradient'' interaction $|\nabla \phi|^2$ in the field theory.  From a computational standpoint, this is very attractive.  The field-theoretic representation is also convenient for analytical studies of field fluctuation effects, such as loop expansions.

A convenient place to begin our discussion of the field theory model is with the mean-field approximation, or equivalently, SCFT.  The mean-field solution neglects all field fluctuations and replaces the sum over field configurations in Eq~\ref{z} by a single most probable configuration -- a saddle point located off the real axis in the complex plane.  For the present model (subject to periodic boundary conditions) the saddle point configuration, obtained from the simultaneous equations $\delta H / \delta w = 0$ and $\delta H / \delta \phi = 0$, is homogeneous (a constant pure imaginary number) in both fields.  Furthermore, due to global charge neutrality, the saddle point solution has a distinctive feature: the mean electrostatic potential is constant so that every positive charge is compensated by an equal negative charge.  It follows that the Coulomb interactions are irrelevant in the mean-field limit and the polyelectrolyte mixture should behave exactly as an analogous mixture of neutral polymers.  This leaves open no possibility for complexation, and indeed we shall see that it is necessary to include fluctuations and account for charge correlations to obtain a coacervate phase.  Thus, the mean-field approximation breaks down completely in the case of a simple mixture of cationic and anionic polyelectrolytes.

Analytically, the next level of sophistication is to account for quadratic field fluctuations about the saddle point solution in the evaluation of Eq~\ref{z}.  This is the leading term in a systematic loop expansion~\cite{amit} known as the \textit{Gaussian} or \textit{one-loop} approximation.  Calculations of this type have been reported previously for models of polyelectrolyte complexation~\cite{borue2,castelnovo1,kudlay1}.  For our symmetric polycation-polyanion mixture, fluctuations of the $w$ and $\phi$ fields \textit{decouple} at the one-loop level.  Moreover, it turns out that the system is parameterized by only \textit{three} dimensionless combinations of variables: reduced polymer concentration $C = 2 n R_g^d / V$, reduced excluded volume parameter $B = u_0 N^2 / R_g^d$, and reduced Bjerrum length $E = K_d l_B \sigma^2 N^2 / R_g^{d-2}$.  Here $R_g = (N b^2 / 2 d)^{1/2}$ is the size of an ideal non-interacting polymer (which coincides with its radius of gyration in 3D), $b$ is the statistical segment length, and $d$ is the number of dimensions.  Thus, the one-loop approximation provides relevant combinations of parameters responsible for the thermodynamic state of the system~\cite{note}.  It should be noted that the electrostatic interactions are manifested only in the parameter $E$.

The one-loop approximation yields analytical expressions for correlation lengths in the polyelectrolyte mixture.  In particular, it predicts that polymer segment density fluctuations are correlated on the scale of the well-known Edwards length~\cite{edwards}.  Of greater interest is the prediction that \textit{charge} density fluctuations are correlated on the scale of the polyelectrolyte length
\begin{equation}
\xi_{PE} = \frac{R_g}{\left( 2 E C \right)^{1/4}} = \left(\frac{b^2}{4 d K_d l_B \sigma^2 \rho}\right)^{1/4},
\label{xi}
\end{equation}
where $\rho = 2 n N / V$ is the segment density.  This correlation length, which has been previously identified in 3D~\cite{castelnovo1,shusharina}, has several important characteristics.  First, it is specific to \textit{charged polymers}, as can be observed from its dependence on the Bjerrum length and the statistical segment length.  Second, $\xi_{PE}$ is independent of the chain length $N$, as is expected for dense polymer systems.  Finally, it is proportional to the $- 1/4$ power of the segment density and hence is \textit{qualitatively} different from the Debye-H\"{u}ckel length $\xi_{DH} = (1 / K_d l_B \sigma^2 \rho)^{1/2}$ for small ions of density $\rho$ carrying charge $\pm \sigma$.  Thus, the attachment of charges to polymer chains creates a coupling between chain conformational statistics and charge density that dramatically changes the electrostatic correlation properties of the solution compared with a conventional small ion electrolyte.

Beyond correlation lengths, the one-loop approximation provides an analytical expression for the Helmholtz free energy that can be used to derive expressions for standard thermodynamic quantities.  For example, the fluctuational electrostatic contribution to the osmotic pressure scales as $- \, A_d (E C)^{d/4} k_B T R_g^{-d} \sim - \, k_B T / \xi_{PE}^d$, where $A_d > 0$ is a known numerical coefficient dependent on dimensionality $d$.  This expression is again qualitatively different from the analogous electrostatic term in Debye-H\"{u}ckel theory for simple electrolytes.  Because the net contribution from charge correlations is \textit{negative}, the one-loop theory can predict polyelectrolyte complexation and the coexistence of dilute and coacervate phases.  The spinodals and binodals for this two-phase region follow directly from the one-loop osmotic pressure expression.  While some of these predictions were obtained previously by slightly different methods~\cite{borue2,castelnovo1,kudlay1}, their validity has been difficult to assess because the next term in the loop expansion (two-loop order) is very tedious to evaluate.

With the advent of the FTS method, we now have a numerical technique that can be used to simulate the full field theory without the assumption of weak charge and density correlations that underpins the loop expansion.  Computer simulations of the full field theory can provide test beds for the analytical loop expansions, just as particle simulation methods complement analytical virial expansions in the particle-based theory.

The advantage (or disadvantage) of FTS over conventional particle-based simulation techniques can be assessed by comparing computational costs.  The cost of particle-based simulations depends on the total number of atomistic or coarse-grained particles in the system.  State-of-the-art particle-based methods require a number of operations per MD step or MC cycle on the order of $n N \ln n N$ for $n$ polyelectrolytes, each with $N$ beads or segments~\cite{limbach}.  On the other hand, the cost of an FTS field update depends on the spatial discretization of the system, rather than the number of particles.  When the computational cell is divided into a lattice of size $M$, each update requires of order $N M \ln M$ operations~\cite{fredrickson}.  In a melt, $M < n$ if the lattice spacing $\Delta x$ in the FTS approach can be taken larger than $V_p^{1/3}$, where $V_p \sim N$ is the volume of a polymer.  This is met, for example, by choosing $\Delta x$ to be a fraction of $R_g \sim N^{1/2}$.  Thus, FTS has an obvious computational advantage for dense systems of long polymers in which a coarse computational grid suffices to capture mesoscopic structure at and beyond the $R_g$ scale.  On the other hand, particle-based methods are advantaged under more dilute conditions, when fewer molecules need to be described on microscopic scales, or when full chemical details are required.  These considerations are rough guides, and further studies are needed to fully elucidate the conditions under which FTS is competitive with existing simulation techniques.

We employ complex Langevin (CL) sampling in our FTS to avoid the sign problem associated with the complex Hamiltonian $H$ and the non-positive definite character of the statistical weight $\exp(- H)$ in the field theory.  This method was originally devised as a strategy for sampling general types of quantum field theories with complex actions~\cite{klauder,parisi} and has been more recently applied in polymer physics~\cite{ganesan,alexander1,alexander2}.  The idea behind this method is to extend the real fields into the complex plane and to compute ensemble averages of observable quantities by sampling fields along a stationary stochastic trajectory in the complex function space.  While this extension to the complex plane doubles the number of field degrees of freedom, the relevant statistical weight becomes a real non-negative distribution of fields $P[W,\Phi]$, where $W \equiv w_R + i w_I$ and $\Phi \equiv \phi_R + i \phi_I$.

A stochastic CL dynamics in the complex function space is employed to generate a Markov sequence of complex fields with stationary distribution $P[W,\Phi]$.  The complex Langevin equations are $\partial W / \partial t = - \lambda \left( \delta H / \delta W \right) + \eta$ and a similar expression for the $\Phi$ field.  Here $W$ and $\delta H / \delta W$ (or $\Phi$ and $\delta H / \delta \Phi$) are \textit{complex}, but the thermal noise $\eta(\mathbf{r},t)$ is a \textit{real} Gaussian white noise, satisfying the usual fluctuation-dissipation theorem with dissipative coefficient $\lambda$.  Since the thermal noise is placed asymmetrically only on the real part of the CL dynamics, the imaginary parts of the equations ensure the sampled fields have nearly constant phases $H_I$ along the Langevin trajectory.  This removes rapid oscillations and improves convergence.  Recent improvements in stochastic integration algorithms for the CL equations have enabled high-resolution, three-dimensional (3D) field-theoretic simulations~\cite{lennon}.

\begin{figure}
\includegraphics{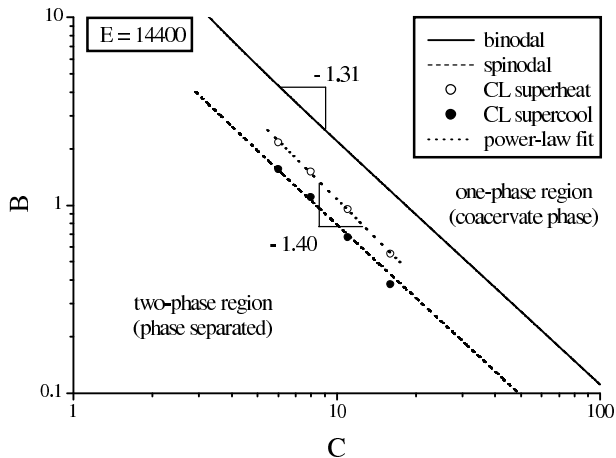}
\caption{\label{two}  Phase diagram for the symmetric polyelectrolyte mixture in 3D plotted in reduced polymer concentration $C$ and reduced excluded volume $B$ at fixed reduced Bjerrum length $E$.  Solid and dashed lines are the analytical one-loop binodal and spinodal, respectively.  Symbols are the result of complex Langevin simulations; the dotted line is a power-law fit.  Numerical simulations were conducted in a cubic cell of size $4 R_g \times 4 R_g \times 4 R_g$ with periodic boundary conditions.}
\end{figure}

Here we report on the application of CL-FTS to polyelectrolyte complexation phenomena, specifically to the same symmetric binary polycation-polyanion model that was used for the analytical one-loop calculations.  A particular focus of our CL-FTS study is the location and size of the two-phase region where coacervation takes place.  An example of such a phase diagram (in 3D) is provided in Fig.~\ref{two}.  Spinodals and binodals for the field theory of Eq~\ref{h} are surfaces in the three-parameter space of the reduced variables $C$, $B$, and $E$.  The figure represents a cross-section of this three-dimensional space by a plane $E = 14400$; hence the diagram involves only the $C$ and $B$ variables.  The diagram features both the CL-FTS results (symbols and dotted-line fit) and the one-loop analytical approximation to the binodal (solid line) and the spinodal (dashed line).  The one-phase region (disordered homogeneous phase) is above and to the right of the lines, and the two-phase region is below and to the left.  The tie lines in the two-phase region are horizontal (constant $B$) and connect nearly pure solvent ($C \approx 0$) with a coacervate phase at the binodal concentration.  The two symbols for the CL simulation data at each concentration $C$ represent the hysteresis upon varying the solvent quality $B$, with the upper symbol corresponding to superheating (increasing $B$) and the lower one corresponding to supercooling (decreasing $B$).

Remarkably, the analytical and numerical results nearly coincide in the high concentration region of the figure, despite the limitations of each method.  The numerical results are subject to finite cell size and chain discretization limitations, while the analytical predictions neglect two-loop and higher order terms in fluctuations.  Nonetheless, our numerical supercooling result practically follows the analytical spinodal, and the analytical binodal yields nearly the same exponent ($- 1.31$) as is obtained from a power-law fit to the numerical superheating points ($- 1.40$).  The cause for the discrepancy between theory and simulation for the location of the binodal is unclear at present, although the overall semi-quantitative agreement indicates that both approaches have utility for this class of problems.

Our future work in polyelectrolyte complexation will extend beyond the symmetric model presented here to include unequal chain lengths, counterions, salt, and explicit solvent.  The latter will allow for the treatment of polyelectrolytes with hydrophobic backbones.  It is important to emphasize that these are straightforward extensions that do not complicate the particle-to-field transformation of the model, nor do they make analytical loop expansions or CL simulations any more difficult.

Another extension is to investigate the effect of \textit{charge distribution} and \textit{polymer architecture} on polyelectrolyte complexation.  This brings us into the realm of block copolyelectrolytes and branched and dendritic polyelectrolyte systems -- all of considerable experimental interest.  As a preliminary example, we have explored a simple variation of our binary polycation-polyanion model where all the charge of each species is concentrated into a small ``charged block'' at the chain end; c.f.\ Fig.~\ref{one}(b).  The remaining longer portion of each chain constitutes a ``neutral block''.  In our simple model, the charged blocks should drive complexation because of the correlation-induced electrostatic attraction, while the neutral blocks repel each other due to the excluded volume interactions in the (assumed) good solvent.  These competing tendencies can drive aggregation to form micellar structures with charged blocks in the micelle cores and the neutral blocks in the coronas.  At relatively low concentrations, the micelles are expected to form a disordered micellar phase, while at higher concentrations they can pack into periodic lattices to form mesophases of various symmetry.  Such \textit{structured coacervate phases} would seem to be of considerable interest for a variety of applications.

\begin{figure}
\includegraphics[scale=0.5]{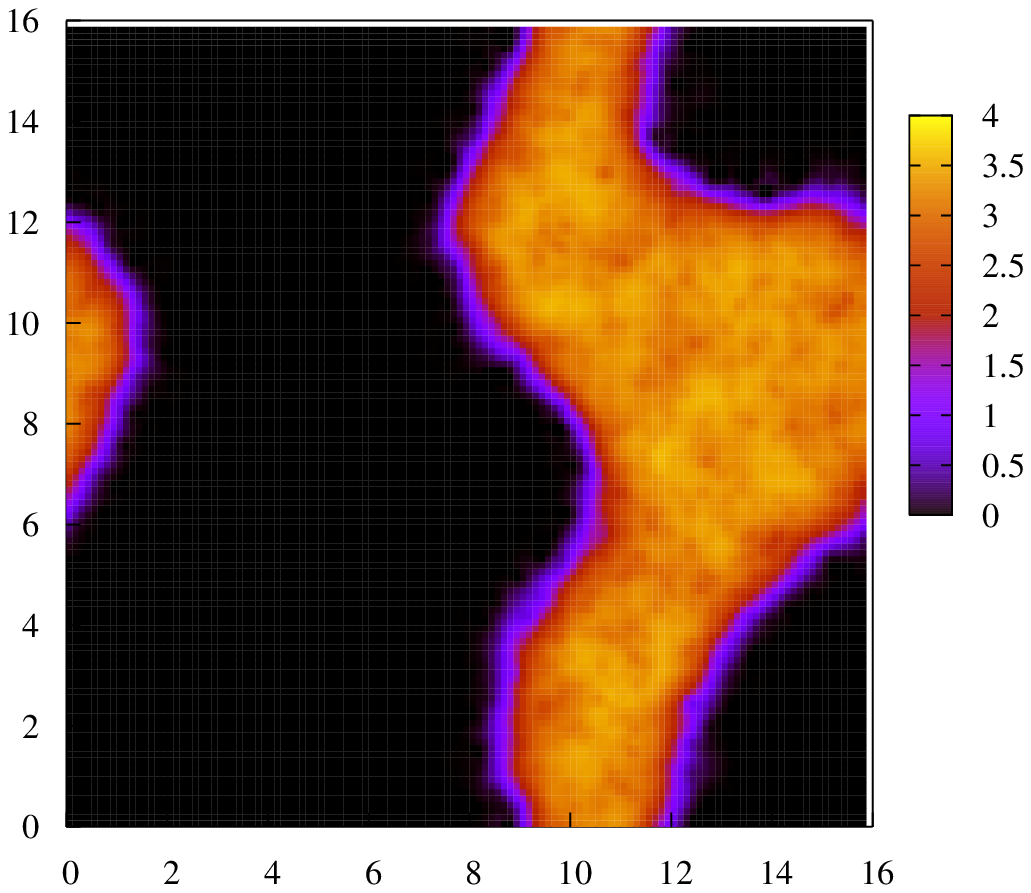}

(a)

\includegraphics[scale=0.5]{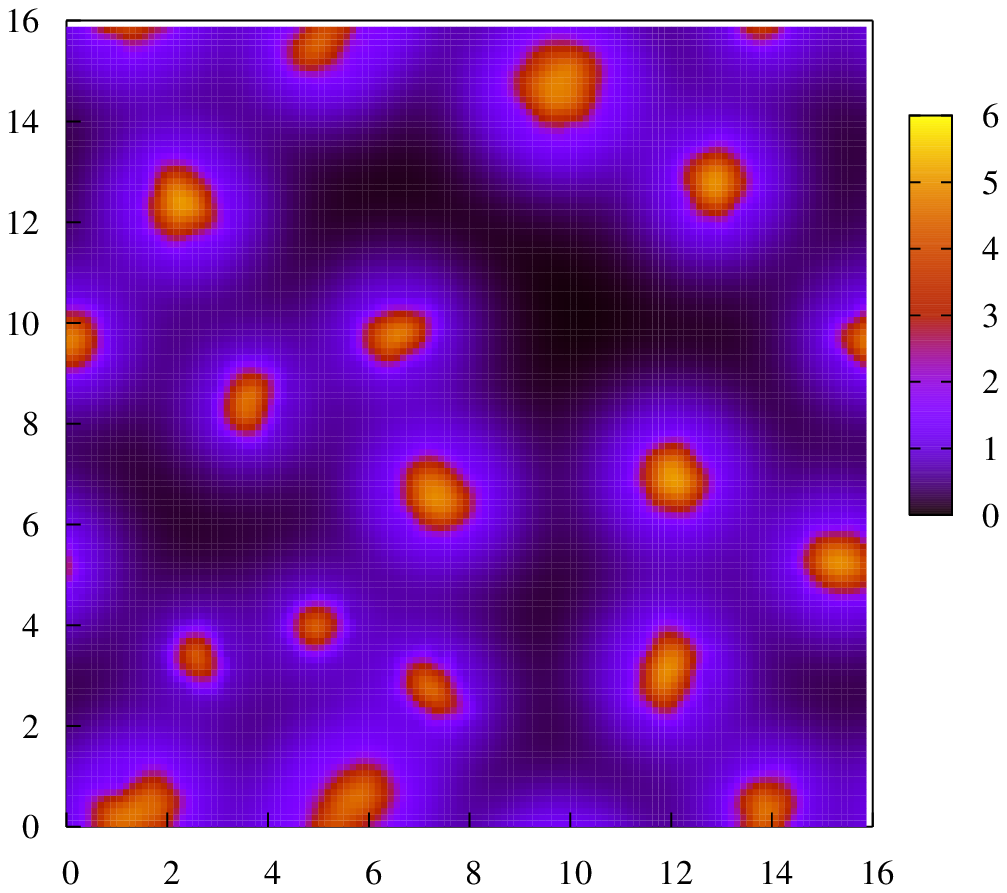}

(b)

\caption{\label{three}  Snapshots of equilibrated field-theoretic simulations.  Normalized total polymer density is presented for the cases of (a) uniformly charged chains and (b) unevenly charged chains, with the same amount of total charge concentrated on 25\% of the chain length.  Macrophases are formed in the uniformly charged case, and a mesoscopic structure develops in the case of block polyelectrolytes.  Both systems are in 2D with periodic boundary conditions.  Cell size is $16 R_g \times 16 R_g$, and the same parameters $C = 6.0$, $B = 0.3$, and $E = 64000$ were used in both cases.}
\end{figure}

In accord with expectations, complex Langevin simulations of the symmetric binary block copolymer model show that the redistribution of charges along the polyelectrolyte chains leads to the formation of aggregates qualitatively similar to micelles.  Figure~\ref{three} illustrates the dramatic difference between the phases formed by uniformly charged chains and unevenly charged chains with the same total charge concentrated into a block of length $N/4$.  The uniformly charged chains [Fig.~\ref{three}(a)] form \textit{macrophases}, i.e.\ large homogeneous phases (a dilute phase and a coacervate phase) constrained only by the size of the computational cell.  On the other hand, the oppositely charged block copolymers [Fig.~\ref{three}(b)] form a \textit{mesophase} with structure on the scale of $R_g$.  The core of each micelle predominantly consists of charged blocks, while the neutral blocks form the repulsive corona.  We are aware of at least one body of experimental work that confirms these (as yet) qualitative predictions~\cite{harada,burgh}.

In summary, field-theoretic simulation methods based on the introduction of auxiliary fields have considerable promise for the investigation of polyelectrolyte complexation phenomena.  These methods possess several distinctive and attractive features.  First, FTS involves exact Hamiltonians and accounts for arbitrarily large fluctuations and strong inhomogeneities.  Thus the technique is not limited in the same ways as the available analytical tools.  Second, field-theoretic methods are especially convenient for treating the long-range Coulomb interaction, which is replaced by a short-ranged square-gradient operator in the auxiliary field representation.  Third, FTS methods are computationally advantaged over particle-based simulations when the systems are dense, polymer chains are long, and the length scale of interest is large (mesoscopic).  Finally, inclusion of additional species (either long chains or small ions) does not lead to a substantial elaboration of the field-theoretic models, nor does it complicate the simulations.

With continued advances in algorithms, we anticipate that FTS methods will enable the study of broad classes of polyelectrolyte systems and phenomena that are beyond the reach of current analytical methods and particle-based computer simulations.  The importance of analytical theory and particle-based simulations will not be diminished, however, by the emergence of FTS techniques.  We see these approaches as complementary, with each contributing valuable insights into the rich structure and thermodynamics of charged polymeric fluids.

\begin{acknowledgments}
The authors are grateful to Fyl Pincus and Kirill Katsov for many valuable discussions and advice.  Acknowledgement is made to the Donors of the American Chemical Society Petroleum Research Fund, the Institute for Collaborative Biotechnology, Rhodia Corporation, and the Mitsubishi Chemical Corporation for the support of this research.
\end{acknowledgments}

\bibliography{references}

\begin{thebibliography}{48}
\bibitem[{not()}]{note}
\bibinfo{note}{It can be shown that the same three parameters $C$, $B$, and $E$
  dictate the thermodynamic properties of the exact field theory, i.e.\ beyond
  the one-loop approximation.}
\end{thebibliography}

\end{document}